\documentclass[letterpaper, 10 pt, conference]{ieeeconf}  % Comment this line out if you need a4paper

\IEEEoverridecommandlockouts                              % This command is only needed if 
                                     
\usepackage{graphicx}
\usepackage{subcaption}
\usepackage{float}
\usepackage{booktabs}
\usepackage{hyperref}
\usepackage{geometry}

\title{\LARGE \bf
Study on the Influence of Embodied Avatars on Gait Parameters in Virtual Environments and Real World*
}

\author{Tianyi Zhou$^{1,2}$, Ding Ding$^{1\dagger}$, Shengyu Wang$^{1}$, Chuhan Shi$^{1}$, Xiangyu Xu$^{1}$% <-this % stops a space
\thanks{*The work is supported by the National Natural Science Foundation of China (62102081)}
\thanks{$\dagger$Corresponding author}% <-this % stops a space
\thanks{$^{1}$School of Computer Science and Engineering, Southeast University, Nanjing, China}%
\thanks{$^{2}$Computer Experimental Teaching Center, Southeast University, Nanjing, China}%
}

\begin{document}

\maketitle
\begin{figure*}[h]
    \centering
    \begin{subfigure}{.33\textwidth} 
        \centering
        \includegraphics[width=\linewidth,height=5cm,keepaspectratio=false]{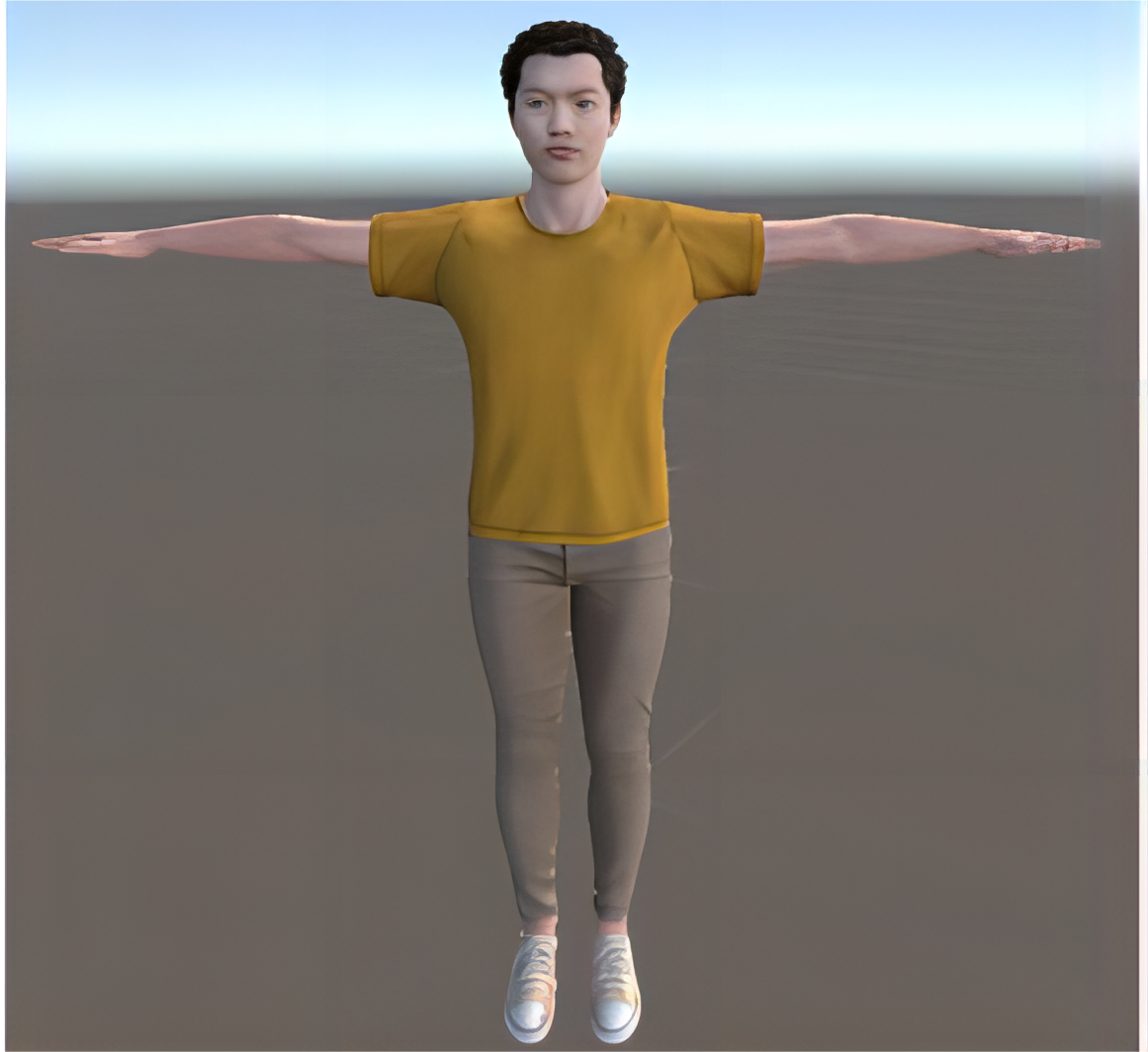} 
        \caption{}
        \label{fig:sub1}
    \end{subfigure}
    \hfill 
    \begin{subfigure}{.33\textwidth} 
        \centering
        \includegraphics[width=\linewidth,height=5cm,keepaspectratio=false]{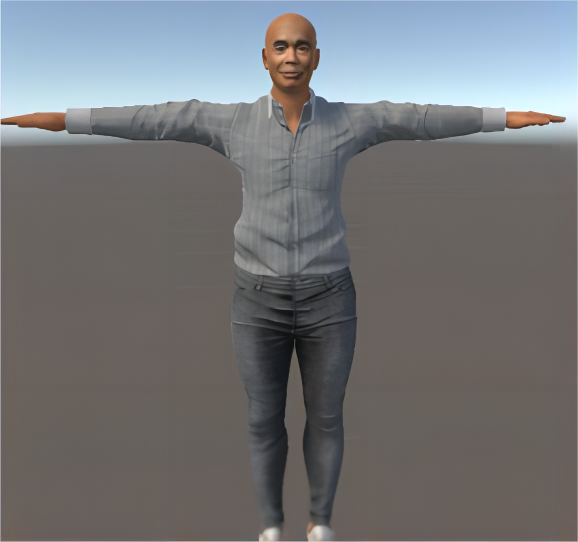} 
        \caption{}
        \label{fig:sub2}
    \end{subfigure}%
    \hfill % 用于分隔子图
    \begin{subfigure}{.31\textwidth} 
        \centering
        \includegraphics[width=\linewidth,height=5cm,keepaspectratio=false]{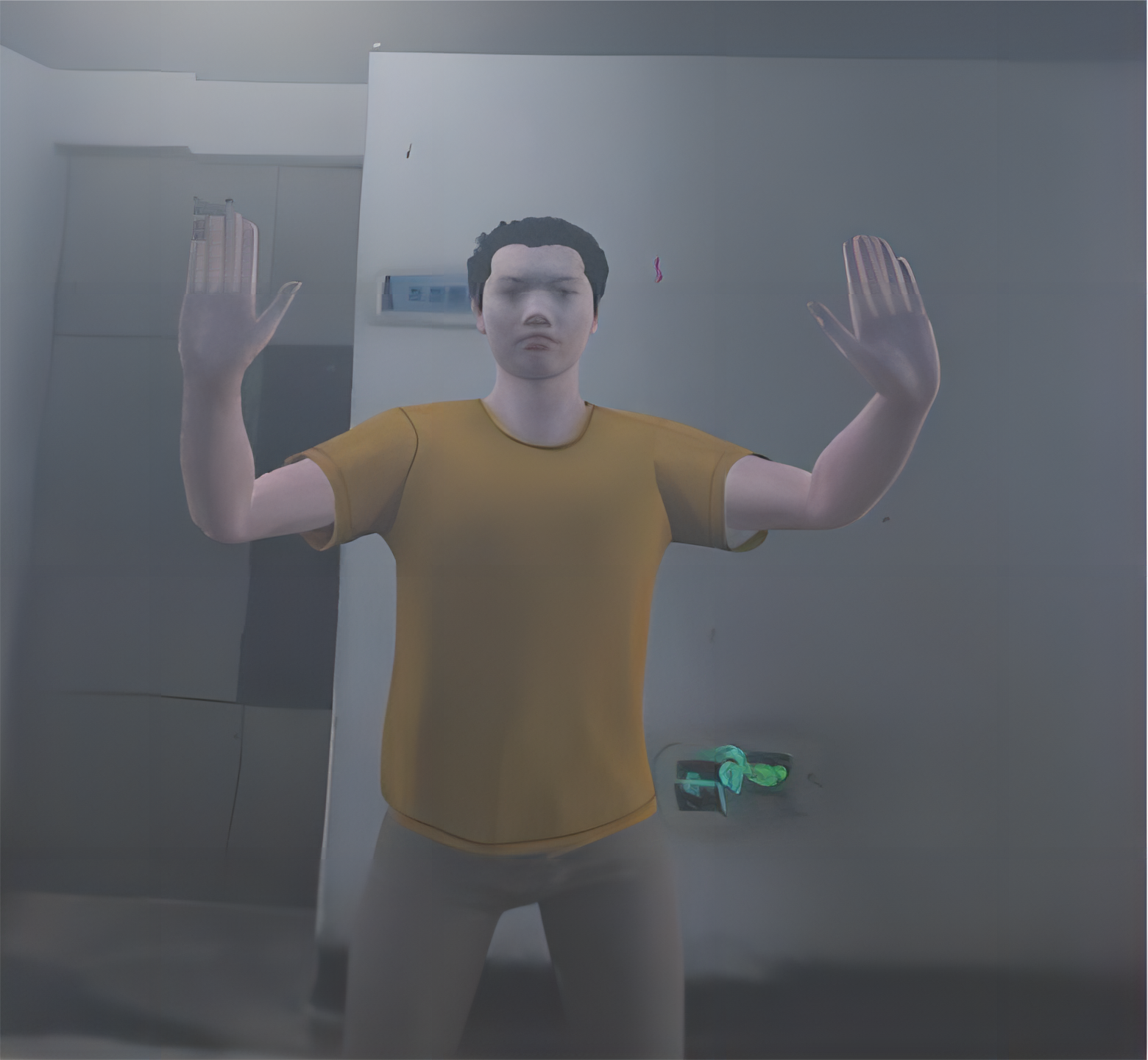} % 调整这里的宽度来适配subfigure宽度
        \caption{}
        \label{fig:sub3}
    \end{subfigure}
    \caption{Avatar created from a user's photo: (a) the same-age avatar, and (b) the old-age avatar. (c) The avatar the user sees in the VR scene.}
    \label{fig1:avatars}
\end{figure*}

\thispagestyle{empty}
\pagestyle{empty}

%%%%%%%%%%%%%%%%%%%%%%%%%%%%%%%%%%%%%%%%%%%%%%%%%%%%%%%%%%%%%%%%%%%%%%%%%%%%%%%%

\begin{abstract}

In this study, we compare the virtual and real gait parameters to investigate the effect of appearances of embodied avatars and virtual reality experience on gait in physical and virtual environments. We developed a virtual environment simulation and gait detection system for analyzing gait. The system transfers real-life scenarios into a realistic presentation in the virtual environment and provides look-alike same-age and old-age avatars for participants. We conducted an empirical study and used subjective questionnaires to evaluate participants' feelings about the virtual reality experience. Also, the paired sample t-test and neural network were implemented to analyze gait differences. The results suggest that there are disparities in gait between virtual and real environments. Also, the appearance of embodied avatars could influence the gait parameters in the virtual environment. Moreover, the experience of embodying old-age avatars affects the gait in the real world.

\end{abstract}


%%%%%%%%%%%%%%%%%%%%%%%%%%%%%%%%%%%%%%%%%%%%%%%%%%%%%%%%%%%%%%%%%%%%%%%%%%%%%%%%
\section{INTRODUCTION}

Gait analysis is an effective clinical method to objectively describe human walking movements and assist in the treatment of diseases such as Parkinson's disease \cite{1Parkinson}, cerebral palsy \cite{2cerebral}, and rheumatoid arthritis \cite{3rheumatoid}. 
%Gait is easily influenced by environments and past incidents \cite{4non-fallers,5randomized}. 
However, it is challenging to simulate various environments or incidents in the physical world. Virtual reality (VR) is a well-established tool for simulating various environments \cite{6Industrial,7paediatric}. Therefore, VR has become popular for gait analysis as it allows for the display of immersive environments to design individual training and treatment programs. Although more and more works analyze gait in VR, it is unclear whether the results of gait analysis in virtual environments (VEs) can be applied to the real world, which affects the effectiveness of VR-based medical diagnostics.

Furthermore, the appearance of the avatars that participants embody has an impact on one's self-perception and may change users’ behavior \cite{8Personalization,9Proteus,10social}. According to the psychological concept known as the Proteus effect, users’ cognition may be influenced by the characteristics of their virtual avatars \cite{12Proteus,13behavior}. A preview study demonstrated that the avatar affects users' gait parameters (e.g. walking speed) and exerts greater influence on people with a clear sense of space \cite{14Acting}. However, most studies have focused on the effects of avatars on gait by constructing parts of the body such as legs or upper bodies \cite{15Comparing,16Supervised}, while ignoring the influence of complete avatars on gait parameters.  Embodying complete avatars gives participants a greater sense of embodiment and ownership than embodying incomplete avatars. It is necessary to investigate the effects of experiencing embodied virtual avatars, especially in the first-person perspective, on the virtual gait. From the first-person perspective,  there are few studies investigating whether synchronizing users' physical and virtual movements and different appearances of complete embodied avatars would affect virtual gait.

Although the effect of avatar embodiment on walking speed in the physical world has been studied by some researchers \cite{11Advantages, 14Acting}, they did not focus on the appearance of avatars. We focus on the appearance and age of complete embodied avatars and investigate whether the participants' gait in the physical world is affected after embodying avatars with a large age difference.

In this study, we propose three research questions to investigate the influence of virtual experiences and embodied avatars on gait parameters in real and virtual environments.

\begin{itemize}
  \item $R1$: If participants embody virtual avatars similar in appearance and age, would their gait in a virtual environment be different from their gait in reality?
  \item $R2$: Is there a significant difference in individuals’ gait in a virtual environment when participants embody complete virtual avatars of the same-age and old-age?
  \item $R3$: After participants who embody old-age avatars experience walking in virtual environments, will their gait in the physical world be affected?
\end{itemize}

In summary, the main contributions of our work are listed as follows:
\begin{enumerate}[]
    \item We developed a virtual environment simulation and gait detection system using VR and motion capture technology. This system can rapidly simulate various physical worlds in VEs and collect participants’ characteristics to create complete avatars that closely resemble the participants.
    \item To increase the sense of embodiment, we customize each participant with two complete avatars that look like them but differ in age. One is a same-age avatar and the other is an old-age avatar\ref{fig1:avatars}.
    \item Subjective and objective measures are used to investigate our research questions. In subjective measurement, questionnaires are used to evaluate participants' feelings about the system and the VR experience. In objective measurement, we build a deep learning model to analyze the gait data. Moreover, the paired sample t-test with a 95\% confidence level is implemented to quantitatively analyze the gait parameters.
\end{enumerate}

% For this research, we developed a virtual environment simulation and gait detection system using VR and motion capture technology. Firstly, this system can rapidly simulate various physical worlds in VEs and collect participants’ characteristics to create complete avatars that closely resemble the participants. Also, our system can capture real-time gait information in both the physical and virtual worlds using a Kinect camera. Then, we built a deep learning model to analyze this data. A key advantage of our system is that when participants embody avatars, their movements and the movements of the virtual avatars are completely synchronized. In our system, participants embody the look-alike virtual avatars from a first-person perspective while walking in VEs. Furthermore, subjective and objective measures were used to investigate our research questions.

\section{RELATED WORK}

\subsection{Theory of Gait}

The human gait cycle consists of two main phases \cite{25parameters}: stance and swing. 
%The stance phase is the period with different movements on one foot: initial contact, loading response, mid-stance, and terminal stance. The swing phase, which constitutes 40\% of the gait cycle, is the period when the foot is airborne, from the moment the toe lifts off the ground until the heel's subsequent contact. 
Analysis of these phases is effective for early intervention and diagnosis of several diseases \cite{46gait,45smartphone}. Some studies have analyzed the spatiotemporal, time-domain, and frequency-domain characteristics of subjects' gait to identify the gait patterns of patients \cite{20Detectin,21Estimating,22extraction}. 
% Wang Y. et al. identified the gait patterns of patients with depression providing a non-invasive solution for the recognition and diagnosis of depression. 
In clinical settings, gait analysis combined with VR technology has been applied to patients' rehabilitation regimens \cite{11Advantages}. Other researchers have designed a series of gait training routines for patients with Parkinson's disease \cite{23training} and vestibular dysfunction \cite{24Vestibular}. These studies also indicate that patients may be more receptive to various VR-based therapies than traditional methods.

\subsection{Avatar's Appearance and Age}

Avatars are a form of self-presentation of a user in a virtual world, and different avatars may contribute to behavioral differences \cite{42environment,43embodied}. Some studies investigated the Proteus effect, indicating that users' self-perception and behavior would be influenced by the characteristics of the avatars \cite{11Advantages,12Proteus}. 

Prior work has shown that age \cite{14Acting}, appearance \cite{10social}, and gender \cite{8Personalization} are the main factors that explain how VR avatars affect the user’s gait. Furthermore, some studies indicated that old-age avatars may lead to slower walking speeds, and consequently, longer walking times \cite{31unconscious,32Priming}. Moreover, Reinhard et al. \cite{14Acting} have investigated the effects of avatar embodiment on walking speeds in the physical world, suggesting that when users experience a strong sense of spatial presence, their behavior is more likely to be influenced after leaving VEs. What's more, Vogel et al. \cite{16Supervised} focused on the movement of the upper parts of the avatars, and Canessa et al. \cite{15Comparing} constructed virtual leg avatars based on the contours of the users’ bodies to study the effect of avatars on the user's gait. Whereas, prior research showed that embodying complete avatars increases users' sense of presence in VR, compared to avatars which have only head and hands visible \cite{10social}. In this work, we consider more about when participants embody complete avatars that look exactly like themselves, whether or not the virtual gait is still significantly different from the real gait, and after embodying complete avatars with large age differences, whether the user's gait in the physical world is affected.

\section{SYSTEM}

%Our system builds unique embodied virtual avatars for participants and can incorporate real-life scenes into the virtual environment realistically. In our system, the synchronization of the embodied avatar and the user's movements is achieved by using an inverse kinematics algorithm that tracks the movements of the head-mounted display and the handheld controllers. Furthermore, the system captures real-time gait information using a Kinect camera to analyze gait differences under different walking conditions.

This system is composed of a virtual avatar module, a virtual environment module, a gait capture module, and a gait data processing and analysis module.

\subsection{Virtual Avatar Module}
The virtual avatar module creates look-alike avatars for participants at their current age and in old age using Character Creator 4\footnote[1]{\url{https://www.reallusion.com/character-creator}}, which is widely used to build facial models from photos of participants. Via this software, various body parameters such as arm length and muscle size are adjusted to closely resemble the participants. 
%The final step is that the system initializes the avatar’s skeletal motion into a t-pose format that is advantageous during the motion capture part and also imports the avatars into Unity3D\footnote[2]{\url{https://unity.com/}} for further texture adjustment to present realistic effects.

\subsection{Virtual Environment Module} 
\begin{figure}[h]
    \centering
    \begin{subfigure}{.2\textwidth} 
        \centering
        \includegraphics[width=\linewidth,height = 3.5cm]{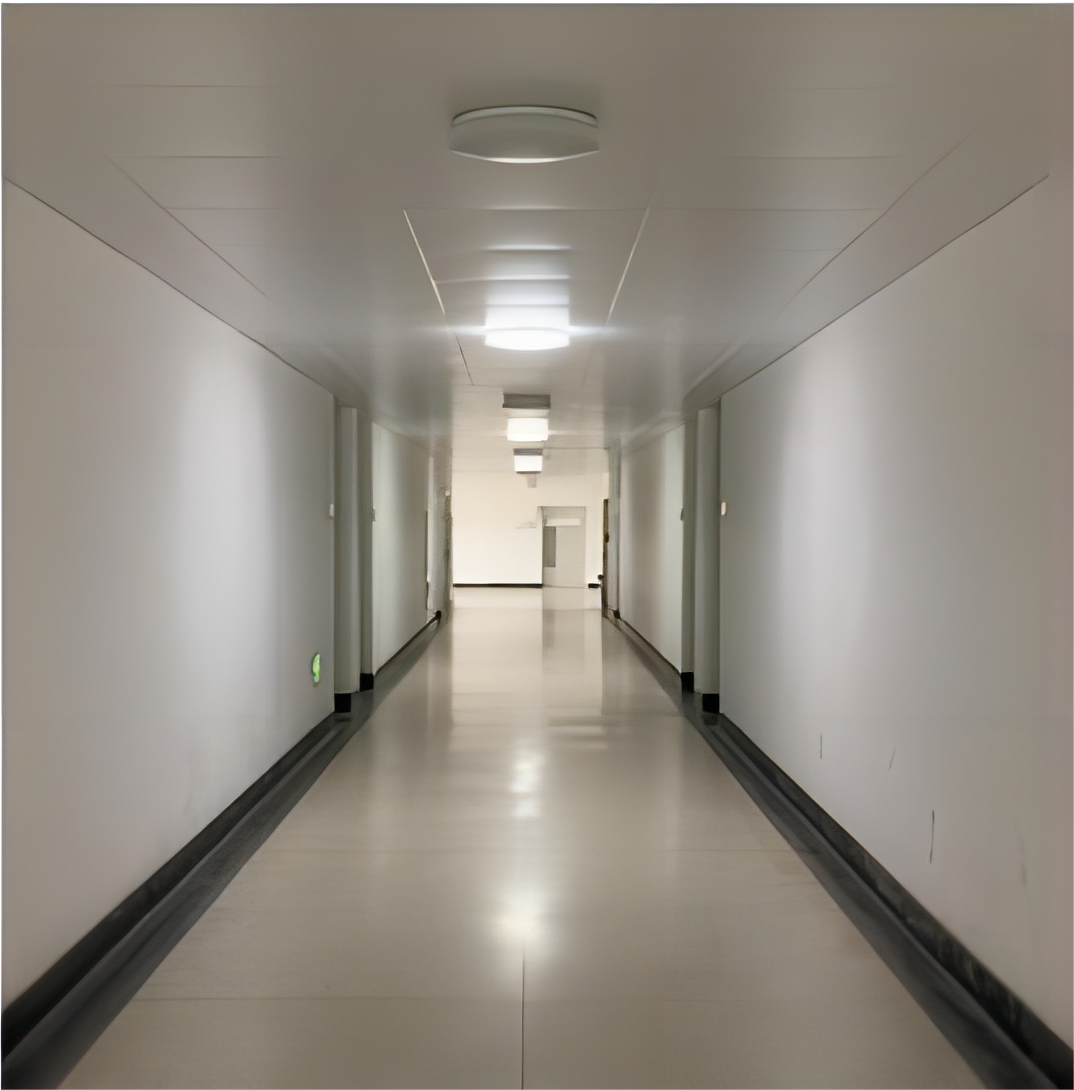} 
        \caption{}
        \label{fig:sub1}
    \end{subfigure}
    
    \begin{subfigure}{.2\textwidth} 
        \centering
        \includegraphics[width=\linewidth,height = 3.5cm]{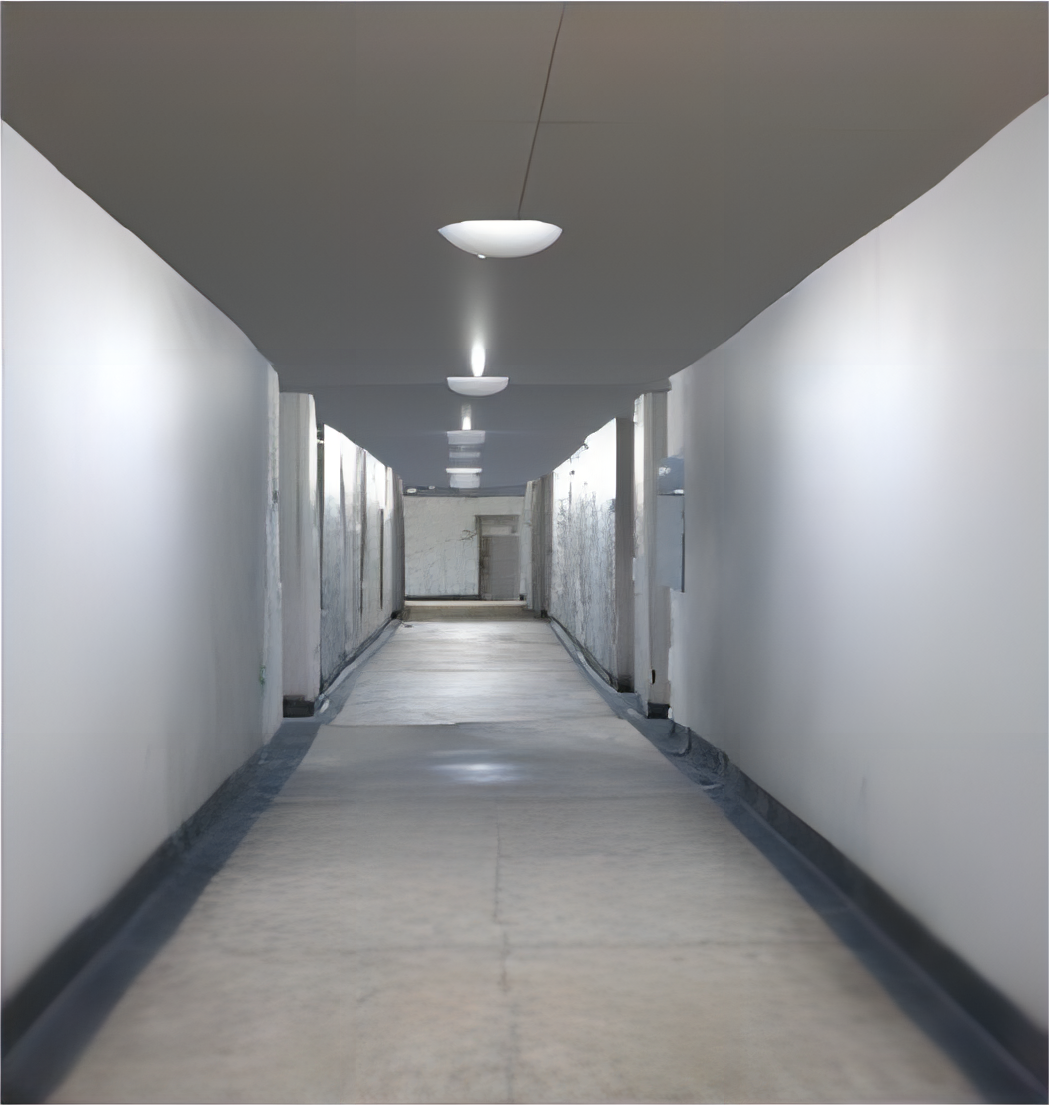} % 调整这里的宽度来适配subfigure宽度
        \caption{}
    \end{subfigure}
    \caption{The walking environments in (a) the real world and (b) the virtual world.}
    \label{corridor}
    
\end{figure}

This system mimics real-world scenes in a virtual environment (Fig. \ref{corridor}). The system created the VR environment using Unity3D and presented it to participants via a head-mounted display. The  sample virtual  scene  was a  5  m $ \times$ 2  m  corridor. The actual corridor was converted into a three-dimensional model using LiDAR technology with the help of modeling software called 3D Scanner. To polish the model and increase the realism of the VE, the system used Blender\footnote[3]{\url{https://www.blender.org/}} to further process the model. To enhance realism, the lighting components in the Unity3D Engine were employed to replicate the lighting conditions found in the actual corridor.

\begin{figure}[h]
  \centering
  \begin{minipage}[b]{0.2\textwidth}
    \includegraphics[height = 4cm]{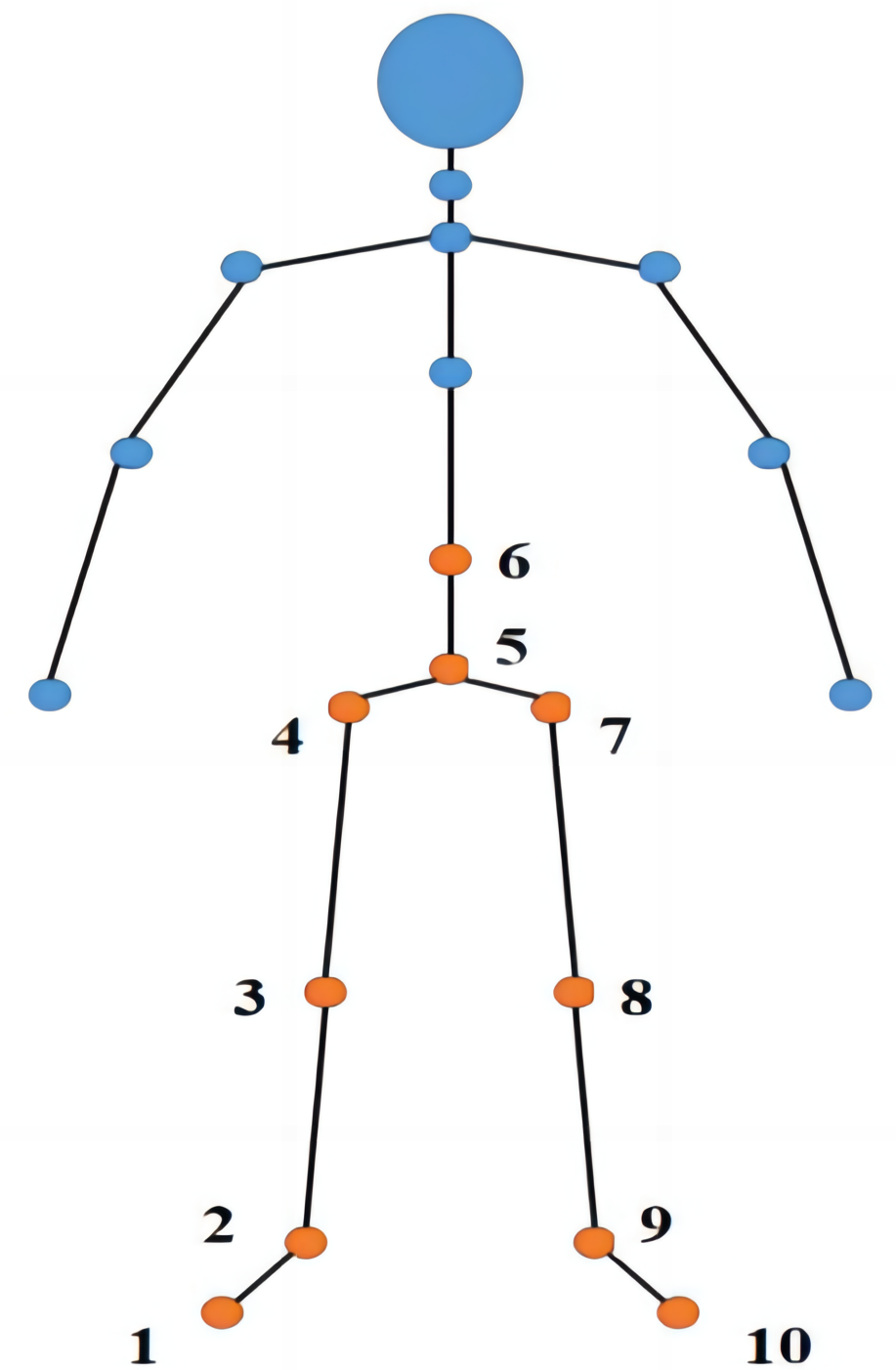}
    \caption*{(a)}
    \label{fig:image1}
  \end{minipage}
  \begin{minipage}[b]{0.2\textwidth}
    \begin{tabular}{c|c}
    \hline
    Number & Name \\ \hline
    1 & FOOT\_RIGHT \\ 
    2 & ANKLE\_RIGHT \\ 
    3 & KNEE\_RIGHT \\ 
    4 & HIP\_RIGHT \\ 
    5 & PELVIS \\ 
    6 & SPINE\_NAVEL \\
    7 & HIP\_LEFT \\ 
    8 & KNEE\_LEFT \\ 
    9 & ANKLE\_LEFT  \\ 
    10 & FOOT\_LEFT \\ \hline
    \end{tabular}
    \caption*{(b)}
    \label{tab:table1}
  \end{minipage}
  \caption{The corresponding names of the captured joint points}
  \label{fig.joint}
\end{figure}

\begin{figure*}[h]
    \centering
    \begin{subfigure}{\textwidth} 
        \centering
        \includegraphics[width=\linewidth,height=6cm,keepaspectratio=false]{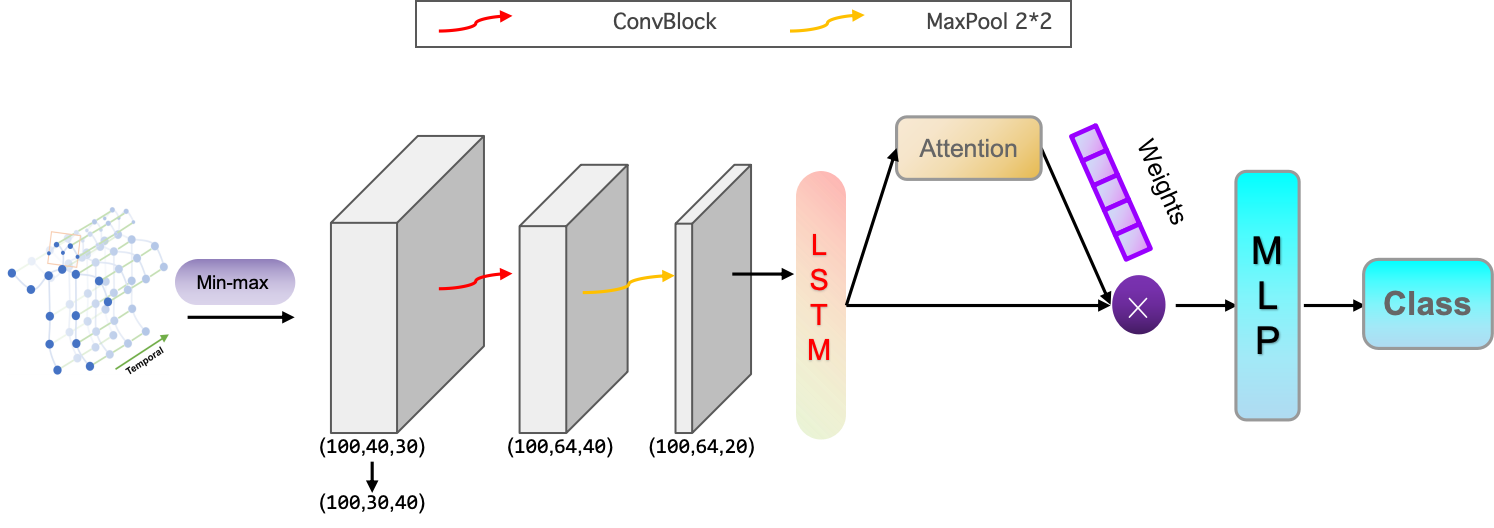} 

    \end{subfigure}
    
    \caption{The architecture of the Neural Network Model}
    \label{fig6}
\end{figure*}

\subsection{Gait Capture Module}
The system uses the position and rotation data from the participants’ heads and both hands to make estimations of the movements in VEs with the Final IK Plugin\footnote[4]{\url{https://assetstore.unity.com/packages/tools/animation/final-ik-14290}}. This approach which uses an inverse kinematics algorithm allows for a rapid and more precise prediction of the positions of other body parts, thereby facilitating the rendering of the user’s virtual avatar in Unity3D and enabling synchronized movement of participants and avatars.

For real-time tracking and recording of human skeletal joint point data, we develop a gait capture program with a Kinect. We mainly pay attention to the gait characteristics so that capture the ten vital joint points of the lower body shown in Fig. \ref{fig.joint}. Meanwhile, the Kinect camera records the participants' skeletal joints as they walk, not only in the real world but also in the virtual environment.

\subsection{Analysis Module}

 The analysis module employs a hybrid structure based on a self-attention mechanism and integrates a convolutional neural network (CNN), a long short-term memory (LSTM). The model classifies participants' gait data under different walking setups and determines whether there is a gait difference based on the accuracy of the classification. As shown in Fig. \ref{fig6}, the CNN initially extracts spatial local features from the sequence of skeletal joint points. Subsequently, the LSTM component processes these features, capturing spatiotemporal information. Additionally, following the LSTM layer, we incorporate an attention layer. This layer plays a crucial role in refining the network’s output by assigning weights to the vectors, thereby identifying which time steps in the sequence hold greater significance for the analysis.

\section{EXPERIMENT}

We designed an empirical study to investigate the effect of embodied avatars on gait parameters in both physical and virtual environments.

\subsection{Participants}

A total of 26 participants (4 female), aged between 18 and 26 years ($M$ = 21.69, $SD$ = 1.59), were involved in the study. These participants are college students from diverse majors and were recruited through verbal invitations and online platforms. No one has physical restrictions or impairments.

\subsection{Materials and Measures}

\subsubsection{Materials} The system captured the skeletal joint points of participants using the Azure Kinect. For the head-mounted display, we used the HTC VIVE Pro headset (2880 $\times$ 1600  pixels, 90 Hz, 110° field of view), while the virtual environments were created in Unity3D. The experiment chose six classical and commonly used gait parameters which are stride length, step length, step width, step height, velocity and cadence for analyzing the gait difference.

\subsubsection{Subjective Measures}To investigate how participants feel about the VR experience, four standardized questionnaires were used. All questionnaires were rated on a 7-point Likert scale from 1 (strongly disagree) to 7 (strongly agree).
\begin{itemize}
  \item Avatar Embodiment. Containing thirteen questions, the questionnaire \cite{36Avatar} comprehensively evaluates the participant’s cognition and experience of virtual avatars from appearance, response ownership, and multisensory.
  \item Igroup Presence Questionnaire. This questionnaire \cite{37Adaptation} is a commonly used presence evaluation tool in VR research, reflecting each participant’s degree of interaction and immersion in virtual environment.
  \item Sense of Agency Scale. Sense of agency \cite{38immediate} is a measure of the degree of conscious control a person has over their mind, body, and immediate environment.                    
  \item Virtual Reality Neuroscience Questionnaire. This questionnaire \cite{39Validation} evaluates the VR-induced symptoms and effects in VR applications.
\end{itemize}

\subsubsection{Objective Measures}The paired sample t-test with a 95\% confidence level was implemented to quantitatively analyze the gait parameters. Furthermore, a deep learning network was implemented to classify participants' gait data under different walking setups and the classification accuracy was used to analyze whether there were differences in gait parameters.

\begin{figure}[h]
    \centering
    
    \includegraphics[width = \linewidth]{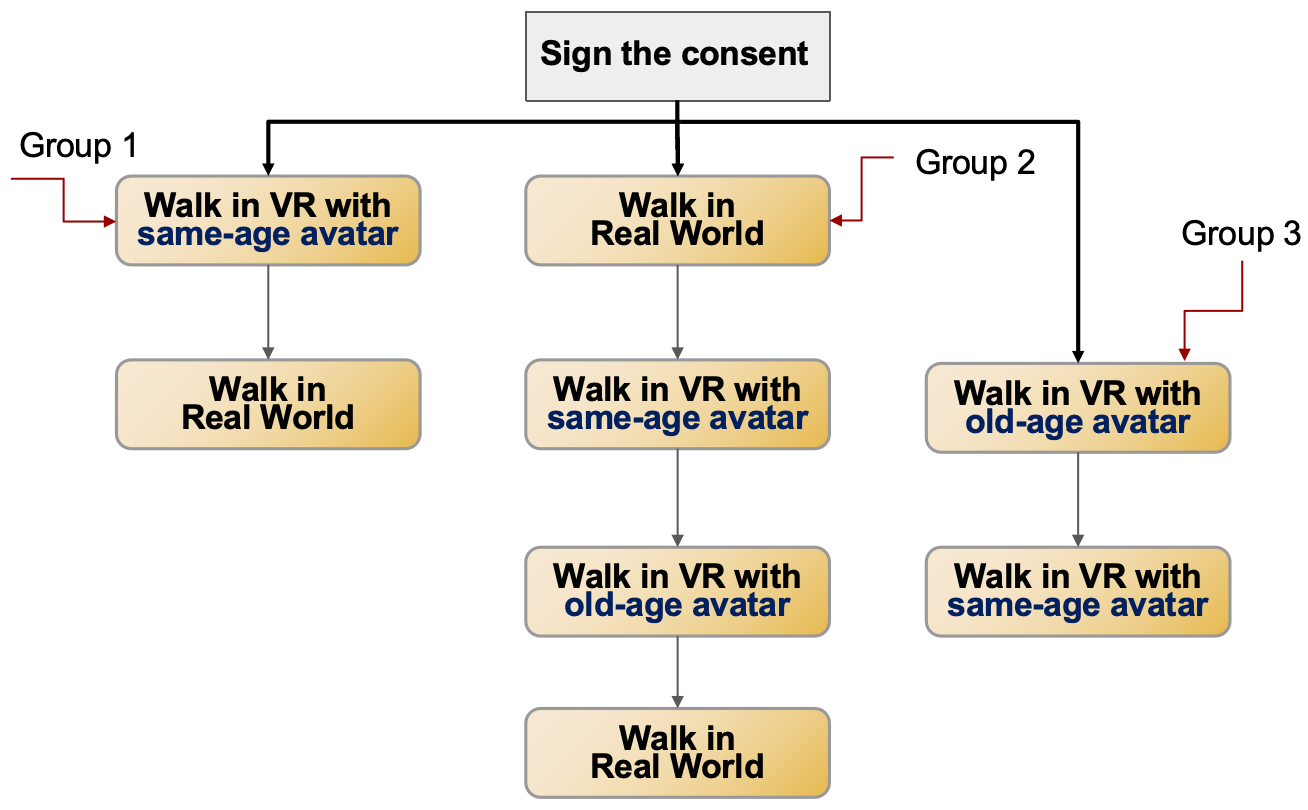} 
 
    \caption{The flow of our experiment}
    \label{fig5}
    
\end{figure}
\subsection{Procedure}

Before the experiment, participants needed to sign their consent to participate, and they were asked to send a photo of themselves to create look-alike same-age and old-age avatars.

The experiment procedure is shown in Fig \ref{fig5}. At the beginning of the experiment, each participant was randomly assigned to one of the three experimental groups. There was a two-minute break between two adjacent parts. In the Group $1$, participants first walked in the virtual environment embodying a same-age avatar. After that, participants were asked to walk back and forth in the real world. As participants in the Group $2$, there were four parts to the experiment. In the first part, participants needed to walk in the real world. Then participants walked in the virtual environment embodying the same-age avatars in the second part and the old-age avatars in the third part. At last, participants were asked to walk in the real world again. In the Group $3$, participants only needed to walk back and forth in the virtual environment. Firstly, they embodied old-age avatars and then embodied same-age avatars.

After participants entered the virtual environment, they first freely observed their avatar's appearance and movements in front of a mirror in a virtual corridor. Participants had 5 minutes to freely observe their avatar's appearance in front of the mirror. In the meantime, they could perform some actions to familiarize themselves with their avatar. After that, they started walking back and forth in the environment according to our commands until the end command was given. Every time they finished the walking part in the virtual environment, we invited participants to complete four questionnaires to measure how they felt about the VR experience. When walking in the real world, participants simply walked back and forth in the real corridor according to the commands.

\begin{table*}

\caption{Paired sample $t$-test for gait parameters in the real world and virtual environments ($M \pm SD $)}
\centering
\begin{tabular}{l c c c c}
\hline
Parameters            & Real World           & Virtual Environment  & $t$    & $p$             \\ \hline
Stride Length (in m)  & 1.241 $\pm$ 0.119    & 1.089 $\pm$ 0.108    & 10.345 & $< 0.001^{**} $ \\
Step Length (in m)    & 0.549 $\pm$ 0.057    & 0.493 $\pm$ 0.047    & 6.304  & $< 0.001^{**} $ \\
Step Height (in m)    & 0.182 $\pm$ 0.028    & 0.164 $\pm$ 0.020    & 2.656  & $0.008^{*}$     \\
Step Width (in m)     & 0.186 $\pm$ 0.032    & 0.203 $\pm$ 0.036    & -2.983 & $0.004^{*}$     \\ 
Cadence (in step/min) & 124.256 $\pm$ 35.237 & 102.879 $\pm$ 23.376 & 2.843  & $0.006^{*}$     \\
Velocity (in m/sec)   & 1.214 $\pm$ 0.460    & 0.856 $\pm$ 0.214    & 4.732  & $< 0.001^{**} $ \\ 
\hline
\multicolumn{5}{l}{Note, the superscript of * means the $p$-value is smaller than 0.05, and ** means the $p$-value is smaller than 0.001. }

\end{tabular}
\label{tab:real and vir}
\end{table*}

\begin{table*}

\caption{Paired sample $t$-test for gait parameters when embodying two avatars ($M \pm SD $)}
\centering
\begin{tabular}{l c c c c}
\hline
Parameters            & Same-age           & Old-age  & $t$    & $p$         \\ \hline
Stride Length (in m)  & 1.061 $\pm$ 0.085    & 1.067 $\pm$ 0.074    & 0.470  & 0.322       \\
Step Length (in m)    & 0.487 $\pm$ 0.030    & 0.486 $\pm$ 0.032    & -0.131 & 0.151       \\
Step Height (in m)    & 0.163 $\pm$ 0.019    & 0.167 $\pm$ 0.023    & 1.067  & 0.449       \\
Step Width (in m)     & 0.202 $\pm$ 0.031    & 0.191 $\pm$ 0.038    & -2.087 & $0.027^{*}$ \\ 
Cadence (in step/min) & 101.794 $\pm$ 21.770 & 102.439 $\pm$ 27.131 & 0.273  & 0.394       \\
Velocity (in m/sec)   & 0.841 $\pm$ 0.223    & 0.830 $\pm$ 0.228    & -0.553 & 0.294       \\ 
\hline
\multicolumn{5}{l}{Note, the superscript of * means the $p$-value is smaller than 0.05, and ** means the $p$-value is smaller than 0.001. }

\end{tabular}
\label{tab:avatars}
\end{table*}

\begin{table*}

\caption{Paired sample $t$-test for gait parameters in the real world before and after embodying the old-age avatar ($M \pm SD $)}
\centering
\begin{tabular}{l c c c c}
\hline
Parameters            & Before           & After  & $t$    & $p$         \\ \hline
Stride Length (in m)  & 1.209 $\pm$ 0.067    & 1.171 $\pm$ 0.078    & 1.454  & 0.092       \\
Step Length (in m)    & 0.529 $\pm$ 0.036    & 0.516 $\pm$ 0.032    & 0.127 & $0.030^{*}$      \\
Step Height (in m)    & 0.172 $\pm$ 0.022    & 0.171 $\pm$ 0.032    & 2.184  & 0.451       \\
Step Width (in m)     & 0.189 $\pm$ 0.038    & 0.188 $\pm$ 0.034    & 0.118 & 0.454 \\ 
Cadence (in step/min) & 101.784 $\pm$ 8.382 & 102.439 $\pm$ 9.714 & 0.496  & 0.317       \\
Velocity (in m/sec)   & 0.929 $\pm$ 0.074    & 0.895 $\pm$ 0.058    & 1.975 & $0.042^{*}$       \\ 
\hline
\multicolumn{5}{l}{Note, the superscript of * means the $p$-value is smaller than 0.05, and ** means the $p$-value is smaller than 0.001. }

\end{tabular}
\label{tab:after}
\end{table*}

\begin{table*}

\caption{Paired sample $t$-test for the four questionnaire scores when embodying different avatars ($M \pm SD $)}
\centering
\begin{tabular}{l c c c c}
\hline
Assessment       & Old-age Avatar                        & Same-age Avatar                   & $t$                        & $p$                       \\  \hline
Appearance       & 4.353 $\pm$ 0.637                     & 4.025 $\pm$ 0.966                     & 2.509                      & $0.023^{*}$               \\
Response         & 4.353 $\pm$ 0.924                     & 4.271 $\pm$ 1.002                     & 0.708                      & 0.489                     \\
Ownership        & 4.412 $\pm$ 1.242                     & 4.694 $\pm$ 1.003                     & -1.021                     & 0.322                     \\
Multisensory     & 4.765 $\pm$ 1.026                     & 4.804 $\pm$ 0.867                     & -0.182                     & 0.858                     \\ 
Embodiment       & 4.471 $\pm$ 0.849                     & 4.448 $\pm$ 0.867                     & 0.17                       & 0.867                     \\
Spatial Presence & 4.929 $\pm$ 0.505                     & 4.729 $\pm$ 0.452                     & 1.047                      & 0.311                     \\ 
Participation    & 4.221 $\pm$ 0.599 & 4.206 $\pm$ 0.510 & 0.124  & 0.903 \\
Realism          & 4.309 $\pm$ 0.682 & 4.647 $\pm$ 0.625 & -2.055 & 0.057 \\
Presence         & 4.567 $\pm$ 0.470  & 4.613 $\pm$ 0.388 & -0.359 & 0.725 \\
Positive Agency  & 4.725 $\pm$ 0.835 & 5.137 $\pm$ 0.710 & -1.641 & 0.120  \\
Negative Agency  & 3.009 $\pm$ 0.712 & 3.109 $\pm$ 0.635 & -0.082 & 0.936 \\
VRISE            & 5.765 $\pm$ 1.297 & 6.047 $\pm$ 1.036 & -1.967 & 0.067 \\ 
\hline
\multicolumn{5}{r}{Note, the superscript of * means the $p$-value is smaller than 0.05, and ** means the $p$-value is smaller than 0.001. }
\end{tabular}
\label{tab:compare}
\end{table*}

\section{RESULTS}

\subsection{Objective Measures}

\subsubsection{Effects of Environment on Gait }
As shown in Table \ref{tab:real and vir}, participants’ stride length ($M$ = 1.089, $SD$ = 0.108), step length ($M$ = 0.493, $SD$ = 0.047), step height ($M$ = 0.164, $SD$ = 0.020) cadence ($M$ = 102.879, $SD$ = 23.376), and velocity ($M$= 0.856, SD = 0.214) were significantly lower in the VEs than in the real world stride length($M$ = 1.241, $SD$ = 0.119), step length ($M$ = 0.549, $SD$ = 0.057), step height ($M$ = 0.182, $SD$ = 0.028), cadence ($M$ = 124.256, $SD$ = 35.237), and velocity  ($M$ = 1.214, $SD$ = 0.460). Whereas in VEs step width ($M$ = 0.203, $SD$ = 0.036) was significantly higher than that ($M$ = 0.186, $SD$ = 0.032) in the real environment.

\subsubsection{Effects of Avatar on Gait}
We compared the two avatars of the same-age and old-age. The results are depicted in Table \ref{tab:avatars}. Participants’ step width when embodying the old-age avatar ($M$ = 0.191, $SD$ = 0.038) was significantly lower ($p$ =0.027) than when embodying the same-age avatar ($M$ = 0.202, $SD$ = 0.031), whereas the other gait characteristics did not show significant differences.

\subsubsection{Effects of Experience on Gait}
% To find whether the virtual experience of embodied avatars would affect the gait parameters in real-world walking, we compared the gait parameters in the real world before and after embodying the elderly avatar. 
As shown in Table \ref{tab:after}, after embodying the old-age avatars, participants’ step length ($M$ = 0.516, $SD$ = 0.032) and velocity ($M$ = 0.895, $SD$ = 0.058) were significantly lower than before.

\subsubsection{Neural network}

\begin{table}[!h]
    \fontsize{10}{13.8}\selectfont%设置字体大小
    \centering
    \caption{Gait classification accuracy}
    \begin{tabular}{c|c}
    \hline
        Condition & Accuracy \\ \hline
        Walking in VE and Real Environment & 91\% \\ 
        Embodying  Different Age Avatars & 52\%\\ 
        Before and After Using the Old-Age Avatar & 50\% \\ \hline
    \end{tabular}
     \label{tab:Network}
\end{table}

The accuracy of the neural network for three different conditions is shown in Table \ref{tab:Network}. For the comparison of gait in the real world and VEs, the model can converge well and classify the two gaits with an accuracy of 91\%.
% However, there is only 52\% accuracy in classifying the gait difference between embodying the same-age and old-age avatars, and 50\% accuracy in classifying the gait in the real world before and after embodying the old-age avatar. 

\subsection{Subjective Measures}

\subsubsection{Avatar Embodiment}
\begin{table}[!h]
    \fontsize{10}{13.8}\selectfont%设置字体大小
    \centering
    \caption{Avatar Embodiment}
    \begin{tabular}{c|c|c|c}
    \hline
        Assessment & $M_{SD}$ & $t$(4) & $p$ \\ \hline
        Appearance  & $4.196_{(0.876)}$ & 1.467 & 0.150 \\ 
        Response  & $4.346_{(1.003)}$ & 2.281 & $0.028^{*}$\\ 
        Ownership  & $4.535_{(1.145)}$ & 3.062 & $0.004^{*}$ \\
         Multisensory  & $4.667_{(1.011)}$ & 4.326 & $< 0.001^{**}$ \\
          Embodiment  & $4.437_{(0.877)}$ & 3.264 & $0.002^{*}$ \\\hline
    \end{tabular}
    \label{tab:Embodiemnt}
\end{table}

As shown in Table \ref{tab:Embodiemnt}, participants' scores for response ($M$ = 4.349, $SD$ = 1.003, $p$ = 0.028), ownership ($M$ = 4.535, $SD$ = 1.145, $p$ = 0.004), multisensory ($M$ = 4.667, $SD$ = 1.011, $p <$ 0.001), and embodiment ($M$ = 4.437, $SD$ = 0.877 , $p$ = 0.002) were significantly higher than the median value ($t$ = 4).

\subsubsection{Igroup Presence Questionnaire}

\begin{table}[!h]
    \fontsize{10}{13.8}\selectfont%设置字体大小
    \centering
    \caption{Igroup Presence Questionnaire}
    \begin{tabular}{c|c|c|c}
    \hline
        Assessment & $M_{SD}$ & $t$(4) & $p$ \\ \hline
         Spatial Presence & $4.856_{(0.489)}$ & 11.486 &$< 0.001^{**}$ \\ 
        Participation  & $4.174_{(0.555)}$ & 2.060 &$0.046^{*}$ \\ 
        Realism  & $4.529_{(0.732)}$ & 4.737 & $< 0.001^{**}$\\ 
          Presence  & $4.601_{(0.444)}$ & 8.887 &$< 0.001^{**}$ \\\hline
    \end{tabular}
    \label{tab:IPQ}
\end{table}

% The participants’ sense of presence was assessed and their mean scores were used as the final score for sense of presence, then subjected to a one-sample t-test with a median value of 4. 
In Table \ref{tab:IPQ}, participants’ scores for spatial presence ($M$ = 4.856, $SD$ = 0.489, $p <$ 0.001), participation ($M$ = 4.174, $SD$ = 0.555, $p$ = 0.046), realism ($M$ = 4.529, $SD$ = 0.732, $p <$ 0.001), and immediacy ($M$ = 4.601, $SD$ = 0.444, $p <$ 0.001) were significantly higher than the median values ($t$ = 4).

\subsubsection{Sense of Agency Scale}
\begin{table}[!h]
    \fontsize{10}{13.8}\selectfont%设置字体大小
    \centering
    \caption{Sense of Agency Scale}
    \begin{tabular}{c|c|c|c}
    \hline
        Assessment & $M_{SD}$ & $t$(4) & $p$ \\ \hline
        Positive   & $5.069_{(0.846)}$ & 8.291 & $<0.001^{**}$ \\ 
        Negative   & $2.947_{(0.738)}$ & -9.353 & $<0.001^{**}$\\ 
       \hline
    \end{tabular}
    \label{tab:Agency}
\end{table}

As shown in Table \ref{tab:Agency}, the scores are significantly higher than the median value ($t$ = 4) for the positive sense of agency ($M$ = 5.069, $SD$ = 0.846, $p <$ 0.001) and significantly lower than for the negative sense of agency ($M$ = 2.947, $SD$ = 0.738, $p <$ 0.001)

\subsubsection{VR-Induced Symptoms and Effects}

\begin{table}[!h]
    \fontsize{10}{13.8}\selectfont%设置字体大小
    \centering
    \caption{VR-Induced Symptoms and Effects}
    \begin{tabular}{c|c|c|c}
    \hline
        Assessment & $M_{SD}$ & $t$(25) & $p$ \\ \hline
        VRISE  & $29.70_{(5.579)}$ & 5.579 & $< 0.001^{**}$ \\ 
       \hline
    \end{tabular}
    \label{tab:Symptoms}
\end{table}

We assessed the symptoms and effects of participants' VR use through the VRISE (VR Induced Symptoms and Effects) section of the VRNQ questionnaire. In Table \ref{tab:Symptoms}, VRISE score ($M$ = 29.70, $SD$ = 5.579, $p <$ 0.001) was significant higher than the value ($t$ = 25).

\subsubsection{Comparison of different embodied avatars}
For participants who embodied different avatars, their scores on the questionnaires after walking were subjected to the paired sample t-test. As shown in Table \ref{tab:compare}, there was a significant difference in the assessment of the appearance ($p$ = 0.023) and the rest of the aspects did not show significant differences.

\section{DISCUSSION AND CONCLUSION}
The result shows the stride length,  step length,  velocity, step height,  and cadence in the virtual environment were significantly lower while step width was significantly higher than the values in the real environment. This finding suggests that there are disparities in gait between virtual and real environments. Meanwhile, since the deep learning model has achieved high accuracy in previous work on gait differences\cite{44spatiotemporal}, our work also indicates the significant gait difference between walking in the real world and VEs by the classification accuracy. When embodying the old-age avatars, the step width is significantly lower than when embodying the same-age avatars. Meanwhile, participants’ assessments of appearance have a significant difference after embodying two avatars. From these results, it can be indicated that the appearance of embodied avatars has an influence on gait in the virtual environment. Moreover, the step length and velocity in the real world after embodying the old-age avatars are significantly lower than the values before embodying suggesting that experiencing old-age avatars affects gait in the real world. What is more, the results of the subjective questionnaires suggest that participants have a good  sense of embodiment, presence, and agency and experience insignificant symptoms and effects via the VR system.

This study had some limitations. Firstly, we only covered samples from university students between the ages of 18 and 26, they are healthy individuals, without gait abnormalities. Secondly, participants only experienced differences in appearance when embodying the old-age avatars but did not experience differences in other kinematic characteristics, such as glaucoma and hunchback.

Some aspects can be improved in the future. 
%For example, we could explore how the findings generalize to larger populations, perhaps using studies outside of a laboratory \cite{41Remote}. 
Since our study only used a sample virtual scene, more experimental scenarios can be added in the future to further investigate the effects of different environmental factors on gait. In the future, we can consider using multiple Kinect cameras to record gait from multiple angles and other sensors, such as gyroscopes and accelerometers, to collect richer and more accurate gait features for future research. In data analysis, the p-values were not adjusted for multiple comparisons, so our work is not very rigorous. Future studies should apply the Bonferroni method for the six gait parameters to ensure more reliable results.

In conclusion, we develop a virtual environment simulation and gait detection system for capturing gait information both in virtual and physical environments and analyzing the gait difference. This system can facilitate future research since it can quickly simulate the real world and easily capture gait data. The result demonstrates that there is a significant difference in gait between the VEs and the real world. Also, the appearance of embodied avatars could influence the gait parameters in the virtual environment. Moreover, the experience of embodying old-age avatars affects the gait in the real world. These findings can provide insight for future studies on gait.

\bibliographystyle{plain}

\bibliography{bb1}

\end{document}